\newcommand{\gray}{\mbox{$\gamma$-ray}}

\newcommand{\pcmq}{\mbox{cm$^{-2}$}}
\newcommand{\psec}{\mbox{s$^{-1}$}}
\newcommand{\pkev}{\mbox{keV$^{-1}$}}
\newcommand{\funit}{\mbox{ph \pcmq \psec}}
\newcommand{\feunit}{\mbox{ph \pcmq \psec \pkev}}
\def\la{\mathrel{\mathchoice {\vcenter{\offinterlineskip\halign{\hfil
$\displaystyle##$\hfil\cr<\cr\sim\cr}}}
{\vcenter{\offinterlineskip\halign{\hfil$\textstyle##$\hfil\cr
<\cr\sim\cr}}}
{\vcenter{\offinterlineskip\halign{\hfil$\scriptstyle##$\hfil\cr
<\cr\sim\cr}}}
{\vcenter{\offinterlineskip\halign{\hfil$\scriptscriptstyle##$\hfil\cr
<\cr\sim\cr}}}}}
\def\ga{\mathrel{\mathchoice {\vcenter{\offinterlineskip\halign{\hfil
$\displaystyle##$\hfil\cr>\cr\sim\cr}}}
{\vcenter{\offinterlineskip\halign{\hfil$\textstyle##$\hfil\cr
>\cr\sim\cr}}}
{\vcenter{\offinterlineskip\halign{\hfil$\scriptstyle##$\hfil\cr
>\cr\sim\cr}}}
{\vcenter{\offinterlineskip\halign{\hfil$\scriptscriptstyle##$\hfil\cr
>\cr\sim\cr}}}}}
\def\MeV{\mbox{Me\hspace{-0.1em}V}}

          [2001/04/25 1.1 (PWD)]
\documentclass[a4paper,twocolumn]{esapub2005} 
\usepackage{times}
\usepackage{natbib}
\usepackage{graphicx}

\title{GRI: The Gamma-Ray Imager mission}
\author{J\"urgen Kn\"odlseder}
\affil{Centre d'Etude Spatiale des Rayonnements, UPS/CNRS, B.P. 44346, 
31028 Toulouse, Cedex, France}
\author{the GRI consortium\footnote{$^2$ the GRI consortium is composed of 
members from the countries 
Belgium,
China,
Denmark, 
France, 
Germany, 
Italy, 
Ireland,
Poland, 
Portugal, 
Russia, 
Spain, 
The Netherlands,
United Kingdom, and
the United States.
A complete list of GRI consortium members can be found on
{\tt http://gri.rm.iasf.cnr.it/}.}}

\begin{document}


\maketitle

\begin{abstract}

With the INTEGRAL observatory, ESA has provided a unique tool to the 
astronomical community revealing 
hundreds of sources, 
new classes of objects, 
extraordinary views of antimatter annihilation in our Galaxy, 
and fingerprints of recent nucleosynthesis processes.
While INTEGRAL provides the global overview over the soft gamma-ray 
sky, there is a growing need to perform deeper, more focused investigations 
of gamma-ray sources. 
In soft X-rays a comparable step was taken going from the Einstein 
and the EXOSAT satellites to the Chandra and XMM/Newton observatories. 
Technological advances in the past years in the domain of gamma-ray 
focusing using Laue diffraction have paved the way towards a new 
gamma-ray mission, providing major improvements regarding sensitivity and 
angular resolution. 
Such a future Gamma-Ray Imager will allow studies of particle acceleration 
processes and explosion physics in unprecedented detail, providing essential
clues on the innermost nature of the most violent and most energetic processes 
in the Universe.
\end{abstract}

\section{From INTEGRAL to GRI}

The present conference has nicely illustrated how INTEGRAL has 
changed our vision of the gamma-ray sky.
The telescopes aboard the satellite have revealed hundreds of sources 
of different types, new classes of objects, extraordinary and 
puzzling views of antimatter annihilation in our Galaxy, and 
fingerprints or recent nucleosynthesis processes.
With the wide fields of view of the IBIS and SPI telescopes, INTEGRAL is 
an exploratory-type mission that allows extensive surveys of the hard 
X-ray and soft gamma-ray sky, providing a census of the source 
populations and first-ever allsky maps in this interesting energy range.
The good health of the instruments after 4 years of operations allows 
the continuation of the exploration during the upcoming years, enabling 
INTEGRAL to provide the most complete and detailed survey ever, 
which will be a landmark for the discipline throughout the next decades.

Based on the INTEGRAL discoveries and achievements, there is now a growing 
need to perform more focused studies of the observed phenomena.
High-sensitivity investigations of point sources, such as compact 
objects, pulsars, and active galactic nuclei, should help to uncover 
their yet poorly understood emission mechanisms.
A deep survey of the galactic bulge region with sufficiently 
high-angular resolution should shed light on the still mysterious 
source of positrons.
And a sensitivity leap in the domain of gamma-ray lines should allow 
the detection of nucleosynthesis products in individual supernova 
events, providing direct insights into the physics of the exploding 
stars.

Technological advances in the past years in the domain of gamma-ray 
focusing using Laue diffraction have paved the way towards a new 
gamma-ray mission that can fulfil these requirements.
Laboratory work and balloon campaigns have provided the 
proof-of-principle for using Laue lenses as focusing devices in  
gamma-ray telescopes \cite{ballmoos04, halloin04}, and concept studies 
by CNES and ESA have demonstrated that such an instrument is 
technically feasible and affordable \cite{duchon06, brown05}.
Complemented by a hard X-ray telescope based on a single-reflection 
multilayer coated concentrator, a broad-band energy coverage can be achieved 
that allows detailed studies of astrophysical sources at 
unprecedented sensitivity and angular resolution, from a few tens 
of keV up to at least 1 \MeV.

Bringing our scientific requirements into the context of these 
technological achievements, we started a common effort to define the
scenario for a future gamma-ray mission that we baptised the 
{\em Gamma-Ray Imager} (GRI).
In this paper we present our scientific motivations, the science 
requirements for the mission, and a mission sketch.
The GRI mission fits well into the framework of ESA's Cosmic Vision 
2015-2025 planning, and it will provide a perfect successor for the 
INTEGRAL mission.
While INTEGRAL provides the general overview over the hard X-ray and 
soft gamma-ray sky, GRI will allow a zoom-in, unveiling the 
physics of cosmic explosions and cosmic accelerators that dominate the 
high-energy Universe.

\section{COSMIC EXPLOSIONS}
\label{sec:explosions}

\subsection{Understanding Type~Ia supernovae}

Although hundreds of Type~Ia supernovae are observed each year, and 
although their optical lightcurves and spectra are studied in great 
detail, the intimate nature of these events is still unknown.
Following common wisdom, Type~Ia supernovae are believed to arise in 
binary systems where matter is accreted from a normal star onto a 
white dwarf.
Once the white dwarf exceeds the Chandrasekhar mass limit a thermonuclear 
runaway occurs that leads to its incineration and disruption.
However, attempts to model the accretion process have difficulties to 
allow for sufficient mass accretion that would push the white dwarf 
over its stability limit \cite{hillebrandt00}.
Even worse, there is no firm clue that Type~Ia progenitors are indeed 
binary systems composed of a white dwarf and a normal star.
Alternatively, the merging of two white dwarfs in a close binary 
system could also explain the observable features of Type~Ia events
\cite{livio03}.
Finally, the explosion mechanism of the white dwarf is only poorly 
understood, principally due to the impossibility of reliably modelling the 
nuclear flame propagation in such objects \cite{hillebrandt00}.

\begin{figure}[!t]
\centering
\includegraphics[width=8cm]{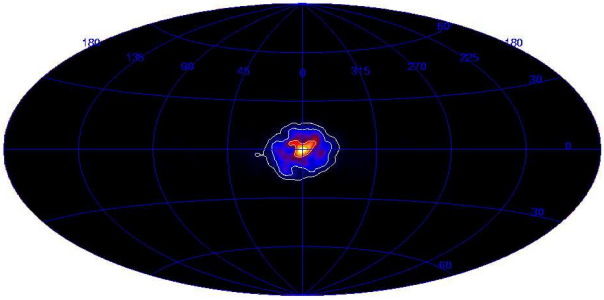}
\caption{
   First all-sky map of 511~keV positron-electron annihilation 
   radiation as observed by the SPI telescope aboard INTEGRAL
   \cite{knoedl05}.
   \label{fig:511keV}}
\end{figure}

In view of all these uncertainties it seems more than surprising that 
Type~Ia are widely considered as standard candles.
Yet, it is this standard candle assumption that is the 
basis of one of the fundamental discoveries of the last decade: 
the accelerating expansion of the Universe \cite{riess98}.
Although empirical corrections to the observed optical lightcurves 
seem to allow for some kind of standardization, there 
is increasing evidence that Type~Ia supernovae are not an homogeneous 
class of objects \cite{mannucci05}.

Gamma-ray observation of Type~Ia supernovae provide a new and unique 
view of these events.
Nucleosynthetic products of the thermonuclear runaway lead to a rich 
spectrum of gamma-ray line and continuum emission that contains a 
wealth of information on the progenitor system, the explosion 
mechanism, the system configuration, and its evolution.
In particular, the radioactive decays of $^{56}$Ni and $^{56}$Co, 
which power the optical lightcurve that is so crucial for the 
cosmological interpretation of distant Type~Ia events, can be 
directly observed in the gamma-ray domain, allowing to pinpoint the 
underlying progenitor and explosion scenario.
The comparison of the gamma-ray to the optical lightcurve 
will provide direct information about energy recycling in the 
supernova envelope that will allow a physical (and not only empirical)
calibration of Type~Ia events as standard candles.

In addition to line intensities and lightcurves, the shapes of the 
gamma-ray lines hold important information about the explosion 
dynamics and the matter stratification in the system.
Measuring the line shapes (and their time evolution) will allow
the distinction between the different explosion scenarios, ultimately 
revealing the mechanism that creates these violent events in the 
Universe \cite{gomez98}.

\subsection{Unveiling the origin of galactic positrons}

The unprecedented imaging and spectroscopy capabilities of the 
spectrometer SPI aboard INTEGRAL have now provided for the first time 
an all-sky image of the distribution of 511~keV positron-electron 
annihilation \cite{knoedl05} (cf.~Fig.~\ref{fig:511keV}).
The outcome of this survey is astonishing: 511~keV line emission is 
primarily seen towards the bulge region of our Galaxy, while the rest of the 
sky remains surprisingly dark.
Only a weak glimmer of 511~keV emission is perceptible from the disk of 
the Galaxy, much less than expected from stellar populations following 
the global mass distribution of the Galaxy.
In other words, positron annihilation seems to be greatly enhanced in 
the bulge with respect to the disk of the Galaxy.

A detailed analysis of the 511~keV line shape measured by SPI has 
also provided interesting insights into the annihilation physics
\cite{churazov05}.
At least two components have been identified, indicating that positron 
annihilation takes place in a partially ionized medium.
This clearly demonstrated that precise 511~keV line shape 
measurements provide important insights into the distribution of the 
various phases of the interstellar medium (ISM) \cite{jean06}.

While INTEGRAL has set the global picture of galactic positron 
annihilation, the source of the positrons still remains
mysterious.
Expected 511~keV line flux levels from individual source candidates 
are (slightly) below the sensitivity of the instruments aboard INTEGRAL.
With its enhanced sensitivity, point-like 511~keV line emission from 
individual objects will get into reach of GRI, enabling the 
measurement of electron-positron annihilation in individual  
candidate sources, such as compact binaries, pulsars, supernova remnants 
or novae.

The discovery of individual positron sources and the measurement of 
their positron production rates would provide a breakthrough in 
gamma-ray astronomy.
The search for the 511~keV line in individual objects is therefore a 
primary objective of the GRI mission.
The measurement of the 511~keV line in individual objects provides also 
an important diagnostics tool, allowing to constrain the physical 
conditions and eventually the plasma composition in the observed 
sources.

\subsection{Understanding core-collapse explosions}

Gamma-ray line and continuum observations address some of the most 
fundamental questions of core-collapse supernovae: 
how and where the large neutrino fluxes couple to the stellar ejecta; 
how asymmetric the explosions are, including whether jets form; 
and what are quantitative nucleosynthesis yields from both static and 
explosive burning processes?

The ejected mass of $^{44}$Ti, which is produced in the innermost ejecta 
and fallback matter that experiences the alpha-rich freezeout of nuclear 
statistical equilibrium, could be measured with GRI to a precision of several 
percent in SN~1987A. 
Along with other isotopic yields already known, this will provide an 
unprecedented constraint on models of that event. 
$^{44}$Ti can also be measured and mapped, in angle and radial velocity, 
in several historical galactic supernova remnants. 
These measurements will help clarify the ejection dynamics, including 
how common jets initiated by the core collapse are.

Wide-field gamma-ray instruments have shown the global diffuse emission 
from long-lived isotopes $^{26}$Al and $^{60}$Fe, illustrating clearly 
ongoing galactic nucleosynthesis. 
A necessary complement to these measurements are high-sensitivity ones of the 
yields of these isotopes from individual supernovae. 
GRI should determine these yields, and map the line emission across several 
nearby supernova remnants, shedding further light on the ejection dynamics. 
It is also likely that the nucleosynthesis of these isotopes in hydrostatic 
burning phases will be revealed by observations of individual nearby massive 
stars with high mass--loss rates.

For rare nearby supernovae, within a few Mpc, we will be given a glimpse 
of nucleosynthesis and dynamics from short-lived isotopes $^{56}$Ni, and 
$^{57}$Ni, as was the case for SN~1987A in the LMC. 
In that event we saw that a few percent of the core radioactivity was 
somehow transported to low-optical depth regions, perhaps surprising 
mostly receding from us, but there could be quite some variety, especially 
if jets or other extensive mixing mechanisms are ubiquitous.

\subsection{Nova nucleosynthesis}

Classical novae are another site of explosive nucleosynthesis that is 
still only partially understood \cite{hernanz05}.
Although observed elemental abundances in nova ejecta are relatively 
well matched by theoretical models, the observed amount of matter 
that is ejected substantially exceeds expectations.
How well do we really understand the physics of classical novae?

Radioactive isotopes that are produced during the nova explosion can 
serve as tracer elements to study these events.
Gamma-ray lines are expected from relatively long living isotopes, 
such as $^{7}$Be and $^{22}$Na, and from positron annihilation of 
$\beta^{+}$-decay positrons arising from the short living $^{13}$N and 
$^{18}$F isotopes.
Observation of the gamma-ray lines that arise from these isotopes
may improve our insight into the physical processes that govern the 
explosion.
In particular, they provide information on the composition of the white 
dwarf outer layers, the mixing of the envelope during the explosion, and 
the nucleosynthetic yields.
Observing a sizeable sample of galactic nova events in gamma-rays should 
considerably improve our understanding of the processes at work, and 
help to better understand the underlying physics.

\section{COSMIC ACCELERATORS}
\label{sec:accelerators}

\subsection{The link between accretion and ejection}

As a general rule, accretion in astrophysical systems is often 
accompanied by mass outflows, which in the high-energy domain take the 
form of (highly) relativistic jets.
Accreting objects are therefore powerful particle accelerators, that 
can manifest on the galactic scale as microquasars, or on the 
cosmological scale, as active galactic nuclei, such as Seyfert 
galaxies and Blazars.

Although the phenomenon is relatively widespread, the jet formation 
process is still poorly understood.
It is still unclear how the energy reservoir of an accreting system 
is transformed in an outflow of relativistic particles.
Jets are not always persistent but often transient phenomena, and it 
is still not known what triggers the sporadic outbursts in accreting 
systems.
Also, the collimation of the jets is poorly understood, and in 
general, the composition of the accelerated particle plasma is not 
known (electron-ion plasma, electron-positron pair plasma).
Finally, the radiation processes that occur in jets are not well 
established.

\begin{figure}[!t]
\centering
\includegraphics[width=8.0cm]{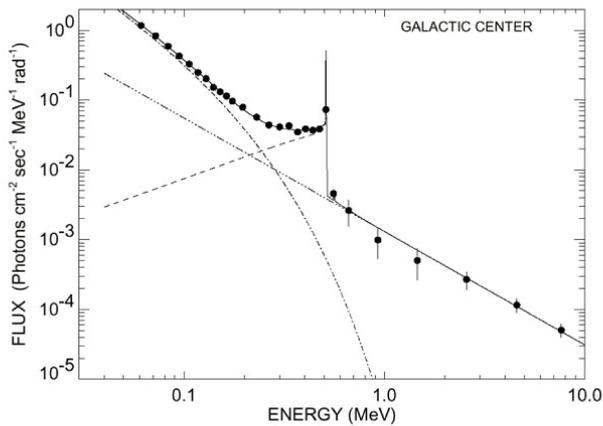}
\caption{
  OSSE hard X-ray and soft $\gamma-$ray spectrum of the Galactic Centre 
  region \cite{kinzer01}. 
  The spectrum is explained by 3 components: 
  an exponentially cut-off powerlaw dominating below $\sim$200~keV, 
  a powerlaw ($\sim E^{-1.7}$) dominating above 511~keV, and 
  a triangular-shaped positronium continuum component plus a narrow 
  line at 511~keV.
  \label{fig:gammasky}}
\end{figure}

Observations in the gamma-ray domain are able to provide a number of 
clues to these questions.
Gamma-rays probe the innermost regions of the accreting systems 
that are not accessible in other wavebands, providing the closest 
view to the accelerating engine.
Time variability and polarization studies provide important insights 
into the physical processes and the geometry that govern the 
acceleration site.
The accelerated plasma may reveal its nature through 
characteristic nuclear and/or annihilation line features which may 
help to settle the question about the nature of the accelerated plasma.

\subsection{The origin of galactic soft \gray\ emission}

For decades, the nature of the galactic hard X-ray ($>$15~keV)
emission has been one of the most challenging mysteries in the field.
The INTEGRAL imager IBIS has now finally solved this puzzle.
Below $\sim$100~keV, about 90\% of the emission has been resolved into 
point sources, settling the debate about the primary origin of the 
emission \cite{lebrun04}.

At higher energies, say above $\sim$200~keV, the situation is less clear.
In this domain, only a small fraction of the galactic emission has so far 
been resolved into point sources, and the nature of the bulk of the 
galactic emission is not entirely explained.
That a new kind of object or emission mechanism should be at work in 
this domain is already suggested by the change of the slope of 
the galactic emission spectrum (cf.~Fig.~\ref{fig:gammasky}).
While below $\sim$200~keV the spectrum can be explained by a 
superposition of Comptonisation spectra from individual point 
sources, the spectrum turns into a powerlaw above this energy, which 
is reminiscent of particle acceleration processes.
Identifying the source of this particle acceleration process, i.e. 
identifying the origin of the galactic soft gamma-ray emission, is 
one of the major goals of GRI.

One of the strategies to resolve this puzzle is to follow the 
successful road shown by INTEGRAL for the hard X-ray emission: 
trying to resolve the emission into individual point sources.
Indeed, a number of galactic sources show powerlaw spectra in the 
gamma-ray band, such as supernova remnants, like the Crab nebula, or 
some of the black-hole binary systems, like Cyg X-1
\cite{mcconnell00}.
Searching for the hard powerlaw emission tails in these objects is 
therefore a key objective for GRI.

\subsection{The origin of the soft \gray\ background}

After the achievements of XMM-Newton and Chandra, the origin of the
cosmic X-ray background (CXB) is now basically solved for energies close 
to a few keV.
Below 8~keV about $\ga$80\% of the emission has been resolved into 
individual sources, which have been identified as active galactic 
nuclei (AGN) \cite{hasinger04}.
Above $\sim$8~keV, however, only $\la$50\% of the CXB has been 
resolved into sources \cite{worsley05}, while in the 20--100~keV 
hard X-ray band, where IBIS is most sensitive, only $\sim$1\% of the 
emission has been resolved \cite{bassani06}.
Above this energy, in the soft \gray\ band, basically nothing is known 
about the nature of the cosmic background radiation.

\begin{figure}[t!]
\centering
\includegraphics[width=8cm]{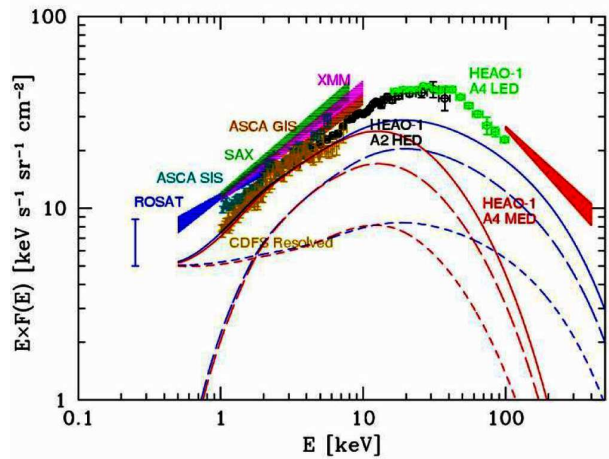}
\caption{
   The 0.25--400~keV cosmic background spectrum fitted with synthesis 
   models \cite{comastri04}. 
   None of the models provides a satisfactory fit of the observations.
   \label{fig:agn}}
\end{figure}

While the situation in the hard X-ray band ($\la$100~keV) may change 
after the launch of the Simbol-X telescope, the soft \gray\ band remains 
unexplored.
It is however this energy band which may provide the key for the 
understanding of the cosmic background radiation.
Synthesis models, which are well established and tested against observational 
results, can be used to evaluate the integrated AGN contribution to the 
soft \gray\ background. 
However, the spectral shape of the different classes of AGN that are 
used for modelling the background has so far not been firmly established 
at soft \gray\ energies. 
As an illustration, Fig.~\ref{fig:agn} shows the impact of the AGN 
power law cut-off energy on the resulting prediction of the cosmic 
background radiation.
Observations by BeppoSAX \cite{risaliti02,perola02} of a 
handful of radio quiet sources, loosely locate this drop-off in the range 
30--300~keV; furthermore these measurements give evidence for a variable 
cut-off energy and suggest that it may increase with 
increasing photon index \cite{perola02}. 
In radio loud sources the situation is even more complicated with some objects
showing a power law break and others no cut-off up to the MeV region. 
In a couple of low luminosity AGN no cut-off is present up to 
300--500~keV. 
The overall picture suggests some link with the absence 
(low energy cut-off) or presence (high energy cut-off)  
of jets in the various AGN types sampled, but the data are still too scarce 
for a good understanding of the processes involved. 

Therefore, a goal of GRI is to measure the soft \gray\ Spectral Energy 
Distribution (SED) in a sizeable fraction of AGN in order to determine 
average shapes in individual classes and so the nature of the radiation 
processes at the heart of all AGN.
This would provide at the same time information for soft \gray\ 
background synthesis models.
On the other hand, sensitive deep field observations should be able to 
resolve the soft \gray\ background into individual sources, allowing 
for the ultimate identification of the origin of the emission.

\subsection{Particle acceleration in extreme B-fields}

\begin{figure}[!t]
\centering
\includegraphics[width=8cm]{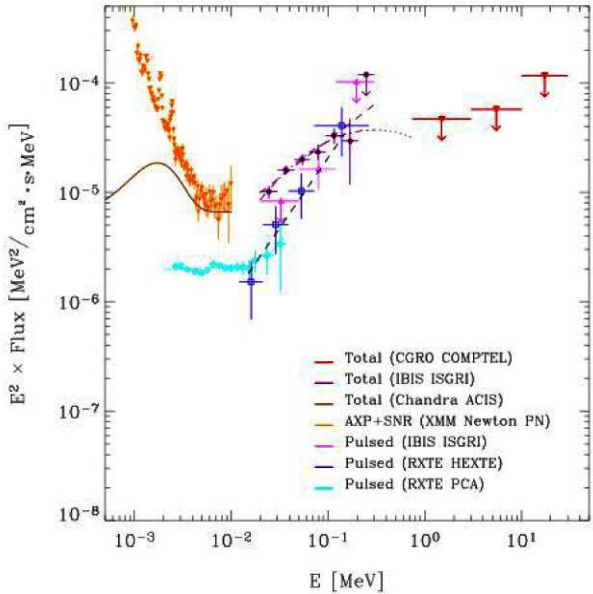}
\caption{
   High-energy emission spectrum from the AXP 1E~1841-045 and the 
   surrounding SNR Kes~73 \cite{kuiper06}.
   The COMPTEL upper limits indicate that the spectra should break 
   between 150--1000~keV.
   \label{fig:axp}}
\end{figure}

The strong magnetic fields that occur at the surface of neutron stars 
in combination with their fast rotation make them powerful 
electrodynamic particle accelerators, which may manifest themselves
to the observer as pulsars.
Gamma-ray emitting pulsars can be divided into 3 classes:
spin-down powered pulsars, 
accretion powered pulsars, 
and magnetically powered pulsars, also known as magnetars.

Despite the longstanding efforts at understanding the physics of 
spin-down powered pulsars, the site of the gamma-ray production within 
the magnetosphere (outer gap or polar cap) and the physical process at 
action (synchrotron emission, curvature radiation, inverse Compton 
scattering) remain undetermined.
Although most of the pulsars are expected to reach their maximum 
luminosity in the MeV domain, the relatively weak photon fluxes have 
only allowed the study of a handful of objects so far.
Increasing the statistics will enable the study of the pulsar
lightcurves over a much broader energy range than today, providing 
crucial clues to the acceleration physics of these objects.

Before the launch of INTEGRAL, the class of anomalous X-ray pulsars 
(AXPs), suggested to form a sub-class of the magnetar population, were 
believed to exhibit very soft X-ray spectra.
This picture, however, changed dramatically with the detection of 
AXPs in the soft gamma-ray band by INTEGRAL \cite{kuiper04,kuiper06}.
In fact, above $\sim$10~keV a dramatic upturn is observed in the 
spectra which is expected to peak in the 100 keV -- 1 MeV domain
(see Fig.~\ref{fig:axp}).
The same is true for Soft Gamma-ray Repeaters (SGRs), as illustrated 
by the recent discovery of quiescent soft gamma-ray emission from 
SGR~1806-20 by INTEGRAL \cite{molkov05}.
The process that gives rise to the observed gamma-ray emission in 
still unknown.
No high-energy cut-off has so far been observed in the spectra, yet 
upper limits in the MeV domain indicate that such a cut-off should be 
present.
Determining what happens in the region of this cut-off may provide 
important insights in the physical nature of the emission process, 
and in particular, about the role of QED effects, such as photon splitting, 
in the extreme magnetic field that occur in such objects.
Strong polarization is expected for the high-energy emission from 
these exotic objects, and polarization measurements may be crucial 
in disentangling the nature of the emission process and the geometry of 
the emitting region.
Measurements of cyclotron features in the spectra will
provide the most direct measure of the magnetic field strengths, 
complementing our knowledge of the physical parameters of the systems.

\subsection{Broad--band gamma-ray emitters}

\begin{figure}[!t]
\centering
\includegraphics[width=8.0cm]{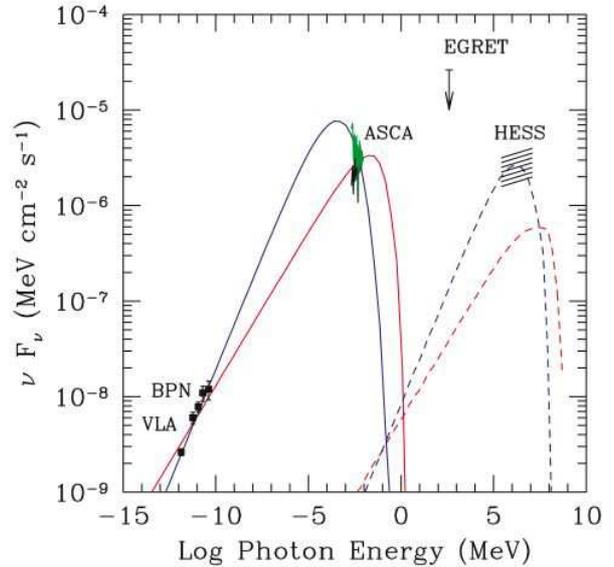}
\caption{
  Broad--band emission spectrum of HESS~J1813-178 assuming that all of the 
  flux originates from the shell of SNR G12.8+0.0.
  Broad--band models based on a primary synchrotron peak and a 
  secondary inverse Compton peak that fit these data points are 
  superimposed \cite{brogan05}.
  \label{fig:hess}}
\end{figure}

The recent advent of the ground-based gamma-ray telescopes HESS and 
MAGIC, operating in the TeV regime, has revealed a substantial number of 
previously unknown high-energy gamma-ray sources in the galactic plane.
The detection of such a new population of galactic GeV--TeV gamma-ray 
emitters has opened a new window for studies of cosmic particle 
accelerators.
Different types of sources have been identified as potential 
counterparts of the high--energy emitters:
isolated pulsars and their pulsar wind nebulae (PWN), 
supernova remnants (SNR), 
star forming regions, 
and binary systems with a collapsed object like a microquasar
or a pulsar. 

For some of the sources observed at TeV energies, hard X-ray and soft 
gamma-ray emission has been detected by INTEGRAL 
\cite{ubertini05,malizia06}, indicating that the emission can by 
explained by a synchrotron inverse Compton mechanism.
Yet, attempts to model the broad-band spectrum have so far been 
unsatisfactory (cf.~Fig.~\ref{fig:hess}) \cite{brogan05,albert06}.
Other sources do not show any hint of radio and/or X-ray emission, 
suggesting that the accelerated particles may be nucleons rather than 
electrons.
Such a conclusion would be challenging, since for the first time, it 
would provide direct evidence for the sources of the main component of 
the cosmic-ray particle spectrum.

It is therefore clear that a better understanding of the emission 
mechanisms of the newly discovered HESS sources requires a 
high-sensitivity broad-band coverage of the entire high-energy 
band, from X-rays over hard X-rays, soft gamma-rays up to the GeV and 
TeV domain.
Observations of TeV sources by GRI (and also of the GeV populations 
that will likely be discovered by AGILE or GLAST), will be 
crucial for the understanding of their emission mechanisms.

\section{THE GRI MISSION}

\subsection{Mission requirements}

Based on our scientific goals, we summarise the GRI mission 
requirements in Table \ref{tab:mission}.
The major requirement for GRI is sensitivity.
Many interesting scientific questions are in a domain where photons 
are rare, and therefore large collecting areas are needed to perform 
measurements in a reasonable amount of 
time\footnote{We baseline as a typical observing time 100~ks}.
It is clear that a significant sensitivity leap is required if the above 
listed scientific questions are to be addressed.

\renewcommand{\footnoterule}{\rule{78mm}{0.1mm}}
\begin{table}
  \begin{center}
    \caption{GRI mission requirements
    (sensitivities are for 100~ks and a detection 
    significance of $3\sigma$).}\vspace{1em}
    \renewcommand{\arraystretch}{1.2}
    \begin{minipage}{\linewidth}
    \begin{tabular}[h]{lll}
      \hline
      Parameter & Requirement & Goal \\
      \hline
      Energy band (keV) & 50 -- 900 & 50 - 1300 \\
      Continuum sensitivity\footnote{units: \feunit}
      & $10^{-7}$ & $3 \times 10^{-8}$ \\
      Narrow line sensitivity\footnote{units: \funit}
      & $3 \times 10^{-6}$ & $10^{-6}$ \\
      Energy resolution & 3\% & 0.5\% \\
      Field of view & 5' & 10' \\
      Angular resolution & 60" & 30" \\
      Time resolution & 100~$\mu$s & 100~$\mu$s \\
      Polarization MDP\footnote{Minimum detectable polarization for 10~mCrab 
      in 100 ks} & 5\% & 1\% \\
      \end{tabular}
      \end{minipage}
    \label{tab:mission}
  \end{center}
\end{table}

With such a sensitivity leap, the expected number of observable sources 
would be large, implying the need for good angular resolution to avoid 
source confusion in crowded regions, such as for example the galactic 
centre.
Also, it is desirable to have an angular resolution comparable to that 
at other wavebands, to allow for source identification and hence 
multi-wavelength studies.

Gamma-ray emission may be substantially polarized due to the non-thermal 
nature of the underlying emission processes.
Studying not only the intensity and the spectrum but also the polarization
of the emission would add a new powerful scientific dimension to the 
observations.
Such measurements would allow the discrimination between the different 
plausible emission processes at work, and would constrain the geometry 
of the emission sites.

\subsection{GRI design}

The key element of GRI is a broad-band gamma-ray lens based on the principle 
of Laue diffraction of photons in mosaic crystals.
Each crystal can be considered as a little mirror which deviates
\gray s through Bragg reflection from the incident beam onto a focal
spot.
Although the Bragg relation 
\begin{equation}
 2 d \sin \theta = n \frac{h c}{E}
 \label{eq:bragg}
\end{equation}
holds only for one specific energy $E$ and its multiples, the mosaic
spread $\Delta \theta$ that occurs in the crystal leads to an energy 
spread
$\Delta E \propto \Delta \theta E^2$
($d$ is the crystal lattice spacing, 
$\theta$ the Bragg angle,
$n$ the diffraction order, 
$h$ the Planck constant, 
$c$ the speed of light and 
$E$ the energy of the incident photon).
Placing the crystals on concentric rings around an optical axis, and
careful selection of the inclination angle for each of the rings,
allows then to build a broad-band gamma-ray lens that has continuous 
energy coverage over a specified band.
Since larger energies $E$ imply smaller diffraction angles $\theta$, 
crystals diffracting large energies are located on the inner rings of the 
lens.
Conversely, smaller energies $E$ are diffracted by crystals located on 
the outer rings.

Several considerations lead us to consider a minimum energy of 
$\sim$200~keV for the Laue lens.
Below this energy, the band pass for individual crystals 
becomes very small, requiring an enormous number of crystal tiles to 
provide a continuum energy coverage.
In addition, machining constraints will probably not allow the use of
crystals that are thinner than $\sim$1--2~mm, hence for energies below 
$\sim$200~keV, absorption of \gray s starts to reduce the efficiency of 
the lens.

The upper energy of the Laue lens is basically set by the focal length 
of the telescope and the smallest radius that can be covered with 
crystal tiles.
Mosaic defocusing, i.e.~the spread of the focused gamma-ray beam due 
to the mosaicity of the crystals, becomes important for focal lengths 
exceeding $\sim$100~metres, reducing the sensitivity gain of the 
instrument.
In addition, for a given energy, the radius on which a given crystal 
has to be placed to focus on the focal spot increases linearly with 
the focal length.
Thus, the minimum energy of the Laue lens drives the total lens 
diameter.
Fixing the minimum energy at $\sim$200~keV and the lens diameter at 
$\la$4~metres, results in a focal length of 60--80~m and a 
maximum energy of $\sim$1~\MeV.

The most promising technology for realizing such a long focal length 
is formation flying of two satellites, one carrying the lens 
and the other the detector.
The focal distance has to be kept to within $\pm10$~cm in order to 
maintain the optimum performances of the instrument.
The size of the focal spot is primarily determined by the size of the 
crystal tiles 
(between $1\times1$~cm$^2$ and $2\times2$~cm$^2$) and 
the mosaic spread $\Delta \theta$ of the crystals (1 arcmin at a 
distance of 100~m corresponds to a size of 3~cm).
Thus the maximum allowed lateral displacement of the detector 
spacecraft with respect to the lens optical axis will be of the order 
of $\pm1$~cm.
Considering the pointing precision, an accuracy of $\sim15$~arcsec 
are sufficient to maintain the system aligned on the source.

Crystals that we currently have under consideration are copper and 
germanium.
Germanium crystals have been employed for the CLAIRE balloon lens, 
which was used to demonstrated the first-ever detection of a gamma-ray 
source by a crystal lens telescope \cite{ballmoos04}.
Copper crystals are currently fabricated at ILL (Grenoble) with the 
required mosaicities, and laboratory measurements indicate that they 
fulfil our efficiency requirements \cite{frontera06}.
We are also studying the possibility of using gradient or bent crystals with 
the aim of substantially increasing the diffraction efficiencies
\cite{barriere06}.
Another possibility is the use of silver or gold crystals, which 
provide good diffraction efficiencies at much less weight than copper.

Although the lens is basically a radiation concentrator (with a beam
size that corresponds to the crystal mosaicity, say $\sim1$ arcmin),
it has a substantial off-axis response.
For sources situated off the optical axis, the focal spot will turn 
into a ring-like structure (which is centred on the lens optical 
axis), with an azimuthal modulation that reflects the azimuthal 
angle of the incident photons.
Thus, the arrival direction of off-axis photons can be reconstructed 
from the distribution of the recorded events on the detector plane.
The field-of-view of the lens is therefore basically restricted by 
the size of the detector.
For a detector size of 30~$\times$~30~cm$^2$ and a focal length of 
100~m the field-of-view amounts to $\sim$15~arcmin.
Within this field-of-view the lens can be used as an (indirect) imaging 
device.
The imaging performances can be considerably improved by employing a 
dithering technique, similar to that employed for INTEGRAL.

It is important to notice that a Laue lens will not significantly 
alter the polarization of the incident radiation.
In other words, a polarized gamma-ray beam will still be polarized 
after concentration on the focal spot, and the use of a polarization 
sensitive detector will allow for polarization measurements.
In view of the expected polarization of non-thermal emission, this 
aspect of GRI opens a new discovery space which will considerably 
improve our understanding of the observed objects.

To profit to the full from the gamma-ray lens, we employ a position 
sensitive detector in the focal spot.
Our actual design studies are mainly focused on a pixelised stack of
detector layers, which on the one hand has the required position 
sensitivity, and on the other hand can be exploited as Compton 
telescope for instrumental background reduction.
Possible detector materials under investigation are CdTe, CZT, Si, 
and/or Ge \cite{caroli06,wunderer06}.
Although Germanium would provide the best energy resolution (and is
certainly the preferred option for detailed studies of gamma-ray 
lines), the related cooling and annealing requirements may drive us 
towards other options.

In order to extend the GRI energy coverage towards energies below 
$\sim$200~keV we plan to add a hard X-ray telescope to the mission.
Such a broad-band coverage is crucial for the understanding of compact 
objects physics, since such sources exhibit generally temporal spectral 
variations over a wide energy band.
In particular, the accurate determination of energy cut-offs will rely 
on an accurate determination of the broad-band spectrum of the object 
under investigation.

The American NuStar mission or the French-Italian Simbol-X mission 
plan to use double-reflection mirrors up to energies of $\sim$80~keV 
to cover the hard X-ray band.
We propose the usage of a single-reflection multilayer-coated 
concentrator to cover the $\sim$50-200~keV energy band, providing 
thus a continuous energy coverage over (at least) the 50-900~keV energy 
band for the GRI mission.
Our model calculations, which are based on measured optical constants, 
predict good effective areas of the concentrator up to $\sim$200~keV and 
even above \cite{christensen06}.
As substrate, we propose the usage of high-precision Si-pore optics 
that are currently developed by ESA in the context of the XEUS 
mission \cite{beijersbergen04}.

\section{CONCLUSIONS}

The gamma-ray band presents a unique astronomical window that allows the 
study of the most energetic and most violent phenomena in our Universe.
With ESA's INTEGRAL observatory, an unprecedented global survey of 
the soft gamma-ray sky is currently performed, revealing hundreds
of sources of different kinds, new classes of objects, extraordinary views 
of antimatter annihilation in our Galaxy, and fingerprints of recent 
nucleosynthesis processes.
While INTEGRAL provides the long awaited global overview over the soft 
gamma-ray sky, there is a growing need to perform deeper, more 
focused investigations of gamma-ray sources, comparable to the 
step that has been taken in X-rays by going from the EINSTEIN  
satellite to the more focused XMM-Newton observatory.
Technological advances in the past years in the domain of gamma-ray 
focusing using Laue diffraction techniques have paved the way towards 
a future gamma-ray mission, that will surpass past missions 
by large factors in sensitivity and angular resolution.
Such a future {\em Gamma-Ray Imager} will allow the study of particle 
acceleration processes and explosion physics in unprecedented depth, 
providing essential clues to the intimate nature of the most violent 
and most energetic processes in the Universe.

\section*{Acknowledgments}

THE GRI project is funded by CNES, ASI and by the Spanish Ministry of 
Education and Science through the project AYA2004-06290-C02-01.


\end{document}